\RequirePackage{lineno}  
\pdfoutput=1 
\documentclass[aps,prl,twocolumn,superscriptaddress,showpacs]{revtex4}

\usepackage{graphicx}  
\usepackage{dcolumn}   
\usepackage{bm}        
\usepackage{xspace}

\usepackage{def}

\newcommand{\zgamma}{$Z/\gamma^{*}$}

\newcommand{\simgt}{\,\hbox{\lower0.6ex\hbox{$\sim$}\llap{\raise0.6ex\hbox{$>$}}}\,}

\newcommand{\simlt}{\,\hbox{\lower0.6ex\hbox{$\sim$}\llap{\raise0.6ex\hbox{$<$}}}\,}

\begin{document} 


%
\title{ Search for a Heavy Top-Like Quark \\
in $p\bar{p}$ Collisions at $\sqrt{s}$ = 1.96 TeV}

\date{\today}
\affiliation{Institute of Physics, Academia Sinica, Taipei, Taiwan 11529, Republic of China} 
\affiliation{Argonne National Laboratory, Argonne, Illinois 60439, USA} 
\affiliation{University of Athens, 157 71 Athens, Greece} 
\affiliation{Institut de Fisica d'Altes Energies, Universitat Autonoma de Barcelona, E-08193, Bellaterra (Barcelona), Spain} 
\affiliation{Baylor University, Waco, Texas 76798, USA} 
\affiliation{Istituto Nazionale di Fisica Nucleare Bologna, $^{bb}$University of Bologna, I-40127 Bologna, Italy} 
\affiliation{Brandeis University, Waltham, Massachusetts 02254, USA} 
\affiliation{University of California, Davis, Davis, California 95616, USA} 
\affiliation{University of California, Los Angeles, Los Angeles, California 90024, USA} 
\affiliation{Instituto de Fisica de Cantabria, CSIC-University of Cantabria, 39005 Santander, Spain} 
\affiliation{Carnegie Mellon University, Pittsburgh, Pennsylvania 15213, USA} 
\affiliation{Enrico Fermi Institute, University of Chicago, Chicago, Illinois 60637, USA}
\affiliation{Comenius University, 842 48 Bratislava, Slovakia; Institute of Experimental Physics, 040 01 Kosice, Slovakia} 
\affiliation{Joint Institute for Nuclear Research, RU-141980 Dubna, Russia} 
\affiliation{Duke University, Durham, North Carolina 27708, USA} 
\affiliation{Fermi National Accelerator Laboratory, Batavia, Illinois 60510, USA} 
\affiliation{University of Florida, Gainesville, Florida 32611, USA} 
\affiliation{Laboratori Nazionali di Frascati, Istituto Nazionale di Fisica Nucleare, I-00044 Frascati, Italy} 
\affiliation{University of Geneva, CH-1211 Geneva 4, Switzerland} 
\affiliation{Glasgow University, Glasgow G12 8QQ, United Kingdom} 
\affiliation{Harvard University, Cambridge, Massachusetts 02138, USA} 
\affiliation{Division of High Energy Physics, Department of Physics, University of Helsinki and Helsinki Institute of Physics, FIN-00014, Helsinki, Finland} 
\affiliation{University of Illinois, Urbana, Illinois 61801, USA} 
\affiliation{The Johns Hopkins University, Baltimore, Maryland 21218, USA} 
\affiliation{Institut f\"{u}r Experimentelle Kernphysik, Karlsruhe Institute of Technology, D-76131 Karlsruhe, Germany} 
\affiliation{Center for High Energy Physics: Kyungpook National University, Daegu 702-701, Korea; Seoul National University, Seoul 151-742, Korea; Sungkyunkwan University, Suwon 440-746, Korea; Korea Institute of Science and Technology Information, Daejeon 305-806, Korea; Chonnam National University, Gwangju 500-757, Korea; Chonbuk National University, Jeonju 561-756, Korea} 
\affiliation{Ernest Orlando Lawrence Berkeley National Laboratory, Berkeley, California 94720, USA} 
\affiliation{University of Liverpool, Liverpool L69 7ZE, United Kingdom} 
\affiliation{University College London, London WC1E 6BT, United Kingdom} 
\affiliation{Centro de Investigaciones Energeticas Medioambientales y Tecnologicas, E-28040 Madrid, Spain} 
\affiliation{Massachusetts Institute of Technology, Cambridge, Massachusetts 02139, USA} 
\affiliation{Institute of Particle Physics: McGill University, Montr\'{e}al, Qu\'{e}bec, Canada H3A~2T8; Simon Fraser University, Burnaby, British Columbia, Canada V5A~1S6; University of Toronto, Toronto, Ontario, Canada M5S~1A7; and TRIUMF, Vancouver, British Columbia, Canada V6T~2A3} 
\affiliation{University of Michigan, Ann Arbor, Michigan 48109, USA} 
\affiliation{Michigan State University, East Lansing, Michigan 48824, USA}
\affiliation{Institution for Theoretical and Experimental Physics, ITEP, Moscow 117259, Russia}
\affiliation{University of New Mexico, Albuquerque, New Mexico 87131, USA} 
\affiliation{Northwestern University, Evanston, Illinois 60208, USA} 
\affiliation{The Ohio State University, Columbus, Ohio 43210, USA} 
\affiliation{Okayama University, Okayama 700-8530, Japan} 
\affiliation{Osaka City University, Osaka 588, Japan} 
\affiliation{University of Oxford, Oxford OX1 3RH, United Kingdom} 
\affiliation{Istituto Nazionale di Fisica Nucleare, Sezione di Padova-Trento, $^{cc}$University of Padova, I-35131 Padova, Italy} 
\affiliation{LPNHE, Universite Pierre et Marie Curie/IN2P3-CNRS, UMR7585, Paris, F-75252 France} 
\affiliation{University of Pennsylvania, Philadelphia, Pennsylvania 19104, USA}
\affiliation{Istituto Nazionale di Fisica Nucleare Pisa, $^{dd}$University of Pisa, $^{ee}$University of Siena and $^{ff}$Scuola Normale Superiore, I-56127 Pisa, Italy} 
\affiliation{University of Pittsburgh, Pittsburgh, Pennsylvania 15260, USA} 
\affiliation{Purdue University, West Lafayette, Indiana 47907, USA} 
\affiliation{University of Rochester, Rochester, New York 14627, USA} 
\affiliation{The Rockefeller University, New York, New York 10065, USA} 
\affiliation{Istituto Nazionale di Fisica Nucleare, Sezione di Roma 1, $^{gg}$Sapienza Universit\`{a} di Roma, I-00185 Roma, Italy} 

\affiliation{Rutgers University, Piscataway, New Jersey 08855, USA} 
\affiliation{Texas A\&M University, College Station, Texas 77843, USA} 
\affiliation{Istituto Nazionale di Fisica Nucleare Trieste/Udine, I-34100 Trieste, $^{hh}$University of Trieste/Udine, I-33100 Udine, Italy} 
\affiliation{University of Tsukuba, Tsukuba, Ibaraki 305, Japan} 
\affiliation{Tufts University, Medford, Massachusetts 02155, USA} 
\affiliation{Waseda University, Tokyo 169, Japan} 
\affiliation{Wayne State University, Detroit, Michigan 48201, USA} 
\affiliation{University of Wisconsin, Madison, Wisconsin 53706, USA} 
\affiliation{Yale University, New Haven, Connecticut 06520, USA} 
\author{T.~Aaltonen}
\affiliation{Division of High Energy Physics, Department of Physics, University of Helsinki and Helsinki Institute of Physics, FIN-00014, Helsinki, Finland}
\author{B.~\'{A}lvarez~Gonz\'{a}lez$^v$}
\affiliation{Instituto de Fisica de Cantabria, CSIC-University of Cantabria, 39005 Santander, Spain}
\author{S.~Amerio}
\affiliation{Istituto Nazionale di Fisica Nucleare, Sezione di Padova-Trento, $^{cc}$University of Padova, I-35131 Padova, Italy} 

\author{D.~Amidei}
\affiliation{University of Michigan, Ann Arbor, Michigan 48109, USA}
\author{A.~Anastassov}
\affiliation{Northwestern University, Evanston, Illinois 60208, USA}
\author{A.~Annovi}
\affiliation{Laboratori Nazionali di Frascati, Istituto Nazionale di Fisica Nucleare, I-00044 Frascati, Italy}
\author{J.~Antos}
\affiliation{Comenius University, 842 48 Bratislava, Slovakia; Institute of Experimental Physics, 040 01 Kosice, Slovakia}
\author{G.~Apollinari}
\affiliation{Fermi National Accelerator Laboratory, Batavia, Illinois 60510, USA}
\author{J.A.~Appel}
\affiliation{Fermi National Accelerator Laboratory, Batavia, Illinois 60510, USA}
\author{A.~Apresyan}
\affiliation{Purdue University, West Lafayette, Indiana 47907, USA}
% ...
\collaboration{CDF Collaboration\footnote{With visitors from $^a$University of Massachusetts Amherst, Amherst, Massachusetts 01003,
$^b$Istituto Nazionale di Fisica Nucleare, Sezione di Cagliari, 09042 Monserrato (Cagliari), Italy,
$^c$University of California Irvine, Irvine, CA  92697, 
$^d$University of California Santa Barbara, Santa Barbara, CA 93106
$^e$University of California Santa Cruz, Santa Cruz, CA  95064,
$^f$CERN,CH-1211 Geneva, Switzerland,
$^g$Cornell University, Ithaca, NY  14853, 
$^h$University of Cyprus, Nicosia CY-1678, Cyprus, 
$^i$University College Dublin, Dublin 4, Ireland,
$^j$University of Fukui, Fukui City, Fukui Prefecture, Japan 910-0017,
$^k$Universidad Iberoamericana, Mexico D.F., Mexico,
$^l$Iowa State University, Ames, IA  50011,
$^m$University of Iowa, Iowa City, IA  52242,
$^n$Kinki University, Higashi-Osaka City, Japan 577-8502,
$^o$Kansas State University, Manhattan, KS 66506,
$^p$University of Manchester, Manchester M13 9PL, England,
$^q$Queen Mary, University of London, London, E1 4NS, England,
$^r$Muons, Inc., Batavia, IL 60510,
$^s$Nagasaki Institute of Applied Science, Nagasaki, Japan, 
$^t$National Research Nuclear University, Moscow, Russia,
$^u$University of Notre Dame, Notre Dame, IN 46556,
$^v$Universidad de Oviedo, E-33007 Oviedo, Spain, 
$^w$Texas Tech University, Lubbock, TX  79609, 
$^x$IFIC(CSIC-Universitat de Valencia), 56071 Valencia, Spain,
$^y$Universidad Tecnica Federico Santa Maria, 110v Valparaiso, Chile,
$^z$University of Virginia, Charlottesville, VA  22906,
$^{aa}$Yarmouk University, Irbid 211-63, Jordan,
$^{ii}$On leave from J.~Stefan Institute, Ljubljana, Slovenia, 
}}
\noaffiliation


\begin{abstract}
We present the results of a search for pair production of a heavy top-like ($\tprime$) quark decaying to $Wq$ final states using data corresponding to an integrated luminosity of 5.6 fb$^{-1}$ collected by the CDF~II detector in $\ppbar$ collisions at $\sqrt{s}$~=~1.96~TeV. We perform parallel searches for $\tprime \to Wb$ and $\tprime \to Wq$ (where $q$ is a generic down-type quark) in events containing a lepton and four or more jets. By performing a fit to the two-dimensional distribution of total transverse energy versus reconstructed $\tprime$ quark mass, we set upper limits on the $\tprimepair$ production cross section and exclude a standard model fourth-generation $\tprime$ quark decaying to $Wb$ ($Wq$) with mass below 358 (340) GeV/$c^2$ at 95\% CL.

\end{abstract}

\pacs{    }

\maketitle

The top quark is one of the most recently discovered particles of the standard model (SM), and since its discovery~\cite{top_cdf,top_d0}, the data collected at the 
Tevatron have been actively used to test the validity of the SM predictions of the top quark's properties.
The top quark is unique because of its large mass of 173.3  $\pm$ 1.1 GeV/$c^2$~\cite{topmass}, which distinguishes it from the 
other fermions of the SM. It is similar in mass to the weak force carriers ($W$ and $Z$) as well as the expected mass 
range for the proposed SM Higgs boson~\cite{LEPEWWG}. 
One of the simplest  
extension of the SM is a fourth
chiral generation of massive fermions. A fourth generation is predicted in a number of 
theories~\cite{democratic,gut} and is compatible with precision electroweak data~\cite{okun, kribs}. 
Furthermore, its existence would allow for a higher Higgs boson mass~\cite{HePolonskySu} and relax the tension between indirect predictions which point to very low masses~\cite{LEPEWWG} and direct searches~\cite{DirectSearch, fourthGenHiggs}.

Fourth generation fermions with masses much higher than current lower bounds~\cite{PDG} would have sizable radiative corrections to the quark scattering amplitude~\cite{WeakUltraHeavy}, so the
masses of heavy top-like ($\tprime$) quark and heavy down-type ($b^\prime$) quarks
should be in the range of a few hundred GeV/$c^2$~\cite{kribs}. These ranges are accessible at the Tevatron collider. In addition, 
a small mass splitting between $\tprime$ and $b^\prime$ is 
preferred, such that $m(b^\prime)+m(W) > m(\tprime)$,  and 
$\tprime$ decays predominantly to $Wq$ (a $W$ boson and a down-type quark 
$q = d, s, b$)~\cite{kribs, frampton, PDG}. Previously published limits have excluded a $b^\prime$ at masses below 372 GeV/$c^2$~\cite{prevBprime} and a $\tprime$ at masses below 285 GeV/$c^2$,
assuming that the $\tprime$ decays to $Wq$~\cite{previousD0}. 

In this Letter we report on a search for a $\tprime$ quark decaying to $Wq$, where $q$ 
can be either a generic down-type quark or specifically a $b$ quark. We analyze a data set of $\ppbar$ collisions at $\sqrt{s} = 1.96$~TeV corresponding 
to an integrated luminosity of 5.6~fb$^{-1}$ collected by the 
Collider Detector at Fermilab (CDF~II) which is described elsewhere~\cite{cdf}. We 
search for pair production of such quarks using events characterized by a high-$\pt$ lepton, large missing 
transverse energy $\met$~\cite{pseudo} and multiple hadronic jets. We assume that the new quark is 
heavier than the top quark and it is produced by strong interaction processes. With respect to~\cite{previous} the 
analysis described herein utilizes a data
sample approximately seven times larger, and adds a parallel search wherein it is assumed that the $\tprime$ decays to $Wb$.

The data events used in the analysis are collected by triggers that identify at least one high-$\pt$ $e$ or $\mu$ 
candidate~\cite{e_mu} or by a trigger requiring $\met$ plus jets~\cite{met_jet}.
Events are retained only if the electron or muon candidate has $\pt \geq 20$ (25 for the $\tprime \to Wq$ search) GeV/c and satisfies the 
typical CDF identification and isolation requirements~\cite{e_mu}. Jets are reconstructed using a fixed cone algorithm of radius 0.4 in 
azimuth $(\phi)$ - pseudorapidity $(\eta)$ space~\cite{pseudo} and their energy is corrected for 
detector effects~\cite{jets}. We require at least four jets with $\et 
\geq 20$ GeV and $|\eta| < 2.0$. 
 Missing transverse energy is reconstructed using fully corrected calorimeter and muon 
 information~\cite{e_mu} and required to have 
magnitude $\geq 20$ GeV. For the $\tprime \to Wb$ search at least one of the jets must be identified as having originated 
from a bottom quark ($b$-tagged) by a secondary vertex tagging algorithm~\cite{secVtx}. 
In order to reduce the contribution of the multijet (QCD) background for the $\tprime \to Wq$ search we make some additional requirements. We ask that at least two of the jets have $\et \geq 25$ GeV, that $M_{T,W} > 20$~GeV/$c^2$ and that \metsig $>  - 0.05 \cdot M_{T,W} + 
3.5$, where $M_{T,W}$ is the transverse leptonically decaying $W$ boson mass, and \metsig~is the $\met$ significance~\cite{metsig}. 

\begin{table}
  \begin{center}
\begin{tabular}{c|c} \hline \hline
\multicolumn{2} {c} {Selection requirements by search} \\ \hline
$\tprime \to Wq$ & $\tprime \to Wb$ \\ \hline
lepton $p_T \geq 25$ GeV/c 					& 	lepton $p_T \geq 20$ GeV/c \\
$\geq 4$ jets with $E_T \geq 20$ GeV			& 	$\geq 4$ jets with $E_T \geq 20$ GeV \\
2 jets with $E_T \geq 25$ GeV			 		&		\\
$\met \geq 20 $GeV 							&	$\met \geq 20 $GeV \\
$M_{T,W} > 20$ GeV/$c^2$					& 	$\geq 1$ jet identified \\
$\metsig > -0.05 \cdot M_{T,W} + 3.5$			&	as coming from a b-jet\\
Requirements on $\Delta\phi$ between 	&	\\
lead jet $E_T$ or lepton $p_T$ and $\met$							&	\\ \hline \hline
\end{tabular}
  \end{center}
  \caption{Summary of selection criteria}
  \label{TAB:selection}
\end{table}

The main contribution to the selected sample of events comes from $\ttbar$ production, which is modeled using the
\textsc{pythia} v6.216 Monte Carlo (MC) generator~\cite{PYTHIA} assuming $m_t = 172.5$~GeV/$c^2$. 
The ~\textsc {alpgen}~\cite{ALPGEN} v2.10 matrix-element generator interfaced to 
\textsc{pythia} v6.325 is used to simulate $W+$jets and \zgamma + 
jets events. The $W + $jets samples are generated separately for $W + \bbbar~+$ jets, $W + \ccbar~+$ jets, $W + c~+$ jets 
and $W~+$ light flavor. 
Other backgrounds include diboson production ($WW, WZ, ZZ$) modeled with ~\textsc{pythia}, single top-quark
production simulated using \textsc{madgraph}+\textsc{pythia}~\cite{PYTHIA,MADGRAPH} and multi-jet 
QCD events modeled using a jet-triggered data 
sample normalized to a background-dominated region at low $\met$. The signal sample of $\tprimepair$ production is generated with \textsc{pythia}. The detector response in all MC samples is 
modeled by a~\textsc{geant3}-based detector simulation~\cite{GEANT}.

When examining control regions for the $\tprime \to Wq$ search, defined by events having less than four jets but passing all 
the other selection criteria, it was observed that the MC under-predicted events in the tails of jet $E_T$ and lepton $p_T$ distributions.
For events with electrons this observed mis-modeling was found in events with a high $E_T$ lead (highest $E_T$) jet or high lepton $p_T$; for  events with muons
the discrepancy was present for high lepton $p_T$. Since for misreconstructed events a correlation 
between the misreconstructed object and the $\met$ is expected, cuts are placed on the $\Delta\phi$ 
between the physics object in question and the $\met$.
For electron events with lead jet $\et \geq 160$ GeV it is required 
that the $\Delta\phi$ between the $\met$ and the lead jet be at least 0.6. For electron 
events with lepton $\pt \geq 120$ GeV/c it is required that the $\Delta\phi$ between the lepton 
and the $\met$ be less than 2.6. For muon events there are two categories: 
muons coming from high-$\pt$ lepton triggers, and muons from triggers based on high $\met$ plus jets. 
For muons in the first category if the lepton $\pt$ is greater than 120 GeV/c 
it is required that the $\Delta\phi$ between the lepton and the $\met$ be less than 2.6. For muons in the second 
category if the lepton $\pt$ is greater than 120 GeV/c it is required that the $\Delta\phi$ between the lepton and the $\met$ 
be between 0.4 and 2.6. These cuts only reduce our signal efficiency by 0.5\% and greatly improve our modeling of the tails of the kinematic distributions. Our selection requirements for both searches are summarized in Table~\ref{TAB:selection}. After all selection and trigger requirements we observe 1,441 (4,390) events for the $\tprime \to Wb~(Wq)$ search.

 The total transverse energy ($\sht$), defined as
\begin{equation}
\sht = \sum_{jets} E_T + E_{T,\ell} + \met,
\label{Ht}
\end{equation}
serves as a good discriminator between standard model and new 
physics processes associated with production of high mass particles. In addition we make use of the assumption that the $
\tprime$ decay chain is identical to the one of the top quark, and reconstruct its mass ($\mrec$) using the standard $\chi^2$-based
 fit of the kinematic properties of final $\tprime$ decay products, the same technique utilized in  top quark mass measurement analyses~\cite{top-mass}. 

We perform the search for a $\tprime$ signal by employing a two-dimensional (2D) binned likelihood fit in both $\sht$ and 
$\mrec$. In order to improve the discrimination between potential $\tprime$ signal and SM backgrounds, 
we split the events into four samples, based on the number of jets (exactly 4 or $\geq 5$), and good or poor mass 
reconstruction $\chi^2$ ($\chi^2 < 8$ and $\chi^2 \geq 8$). The sample with exactly 4 jets and good $\chi^2$ has the 
largest sample size due to the fact that the majority of $\ttbar$ events (61\% [65\%] out of all $\geq 4$ jet $\ttbar$ events when [not] 
requiring a jet tagged as  a $b$ quark) fall into this category. The $\tprime$ mass reconstruction is best in this category but the $\tprimepair$ 
events are distributed more uniformly than $\ttbar$ events among all four categories of events. 
To ensure sufficient MC statistics on the high energy tails, we developed an algorithm that merges bins with low MC statistics together into super-bins. The super-bins are defined by the requirement that each super-bin in a template has a relative uncertainty due to MC statistics below 40\%.

The fit is conducted simultaneously for four different sets of templates. The likelihood is defined as the product of the Poisson 
probabilities for observing $n_{i,k}$ events in the bin $i,k$ of ($\sht,\mrec$).
The expected number of events in each bin, $\mu_{i,k}$, is given at base by
the sum over all sources indexed by $j$:
\begin{eqnarray}
     \mu_{i,k} = \sum_j L_j \sigma_j \epsilon_{ikj}  \ \ \ .
\label{mu}
\end{eqnarray}
Here the $L_j$ are the integrated luminosities, the $\sigma_j$ are the
cross sections, and the $\epsilon_{ikj}$ are the efficiencies per
bin of ($\sht,\mrec$). We calculate the likelihood as a function of
the $\tprimepair$ cross section, and 
apply Bayes' theorem with a uniform prior in $\sigma$ to obtain a 95\% CL  upper limit or measure the production rate of $\tprimepair$ events.

The production rates for $\tprimepair$ events, $W$ + jets in the 4-jet bins, and $W$ + jets events in the $\geq 5$ jet bins 
are three unconstrained independent parameters in the fit. Production rates for $\ttbar$, single top, dibosons and $Z + $jets~\cite{topxsec,other_xsec1,other_xsec2} are constrained to their theoretically predicted values and uncertainties.
We consider systematic uncertainties that affect only the normalization as well as those affecting 
the normalization and shape 
of the distributions. The normalization uncertainties and their magnitudes are: integrated luminosity (5.6\%), lepton ID scale factors (1\%), 
uncertainty on the parton distribution functions (1\%) and wholly correlated theory uncertainty on the $\tprime$~\cite{MMangano} and $\ttbar$~
\cite{topxsec} cross section (10\%). The shape and normalization systematics and their impact on the expected limit at a $\tprime$ mass of 
360 GeV/$c^2$ (near the observed limit) are : jet energy scale (2.5\%), the $Q^2$ scale at which $W + $ jets MC events are generated (2.5\%), initial and final state 
radiation (2.5\%) and, for the $\tprime \to Wb$ search only, uncertainty on the $b$-tagging of jets ($<$2.5\%). All of the sources of systematic 
errors are treated in the likelihood as nuisance parameters constrained within their expected distributions. We adopt the profiling method ~
\cite{profile} for dealing with these parameters, i.e. the likelihood is maximized with respect to the nuisance parameters. For normalization 
and shape uncertainties we use a vertical morphing technique~\cite{profile} to change both shape and normalization when fitting. For these 
parameters we interpolate quadratically for less than one $\sigma$ variance and extrapolate linearly for beyond one $\sigma$ variance in 
the expectation value.
Taking this into account the likelihood takes the following expression:  
\begin{eqnarray}
    {\cal L}(\sigma_{\tprimepair}|n_{i,k}) = \prod_{i,k,m,j} P(n_{i,k}|\mu_{i,k}) 
           \times  G(\nu_m |\tilde{\nu}_m,\sigma_{\nu_m}) \ \\ \nonumber 
            \times f_X(\nu_j | \tilde{\nu}_j, \sigma_{\nu_j})\ \ \ 
\label{systlikelihood}
\end{eqnarray}
where $\nu_m$ are the nuisance parameters used in the morphing parameters (constrained by gaussian $G$ terms to their 
expectation) and $\nu_j$ are the nuisance parameters used in non-morphing parameters (constrained by log normal $f_X$
terms to their expectations), such as $\sigma_{\ttbar}$, 
$L_j$ and etc. ${\tilde{\nu}}_{m,j}$ are their central nominal values
and $\sigma_{\nu_{m,j}}$ are their uncertainties.

\begin{figure}[htbp]
  \begin{center}
   \includegraphics[width=8.5cm,clip=]{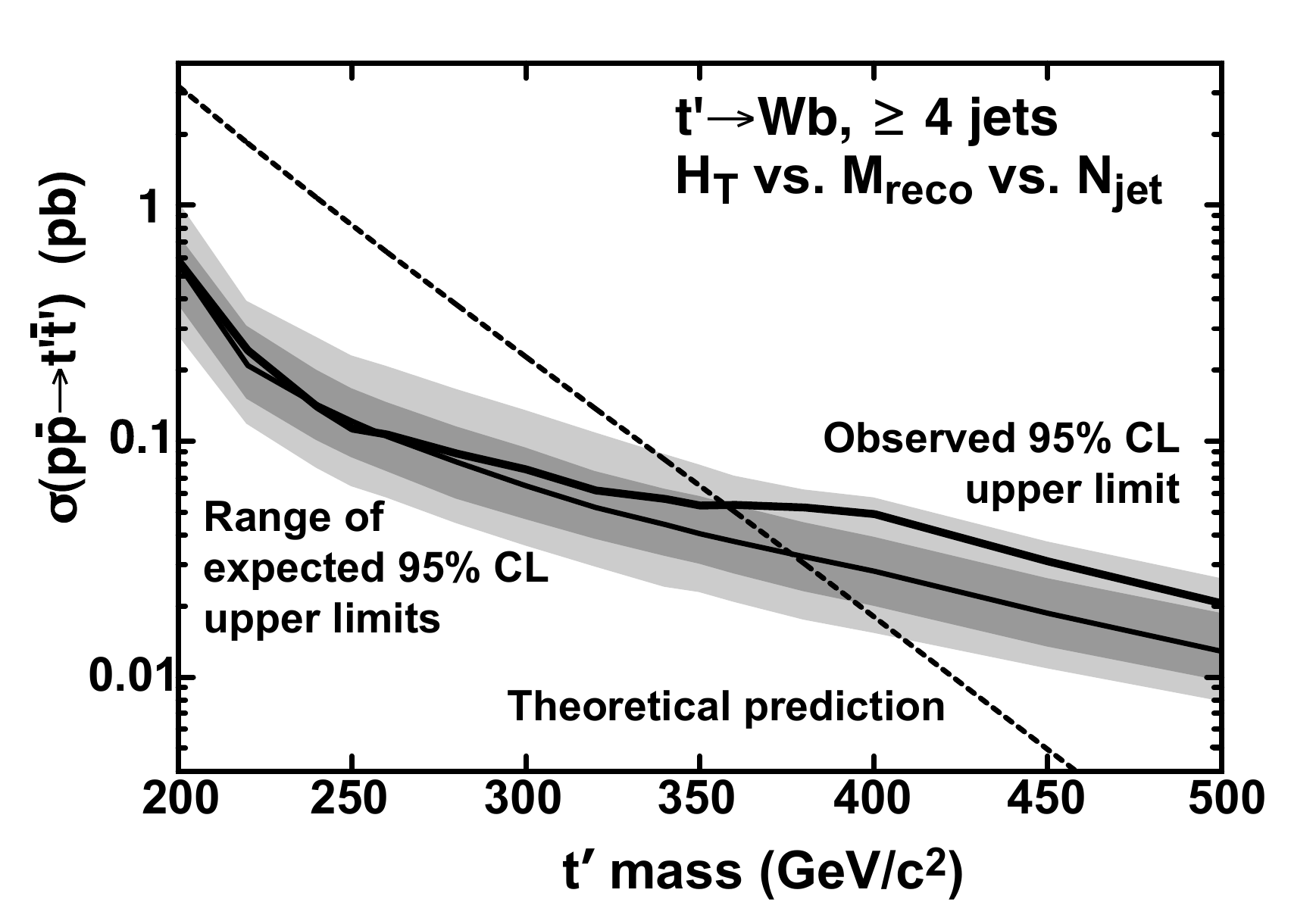}
         \includegraphics[width=8.5cm,clip=]{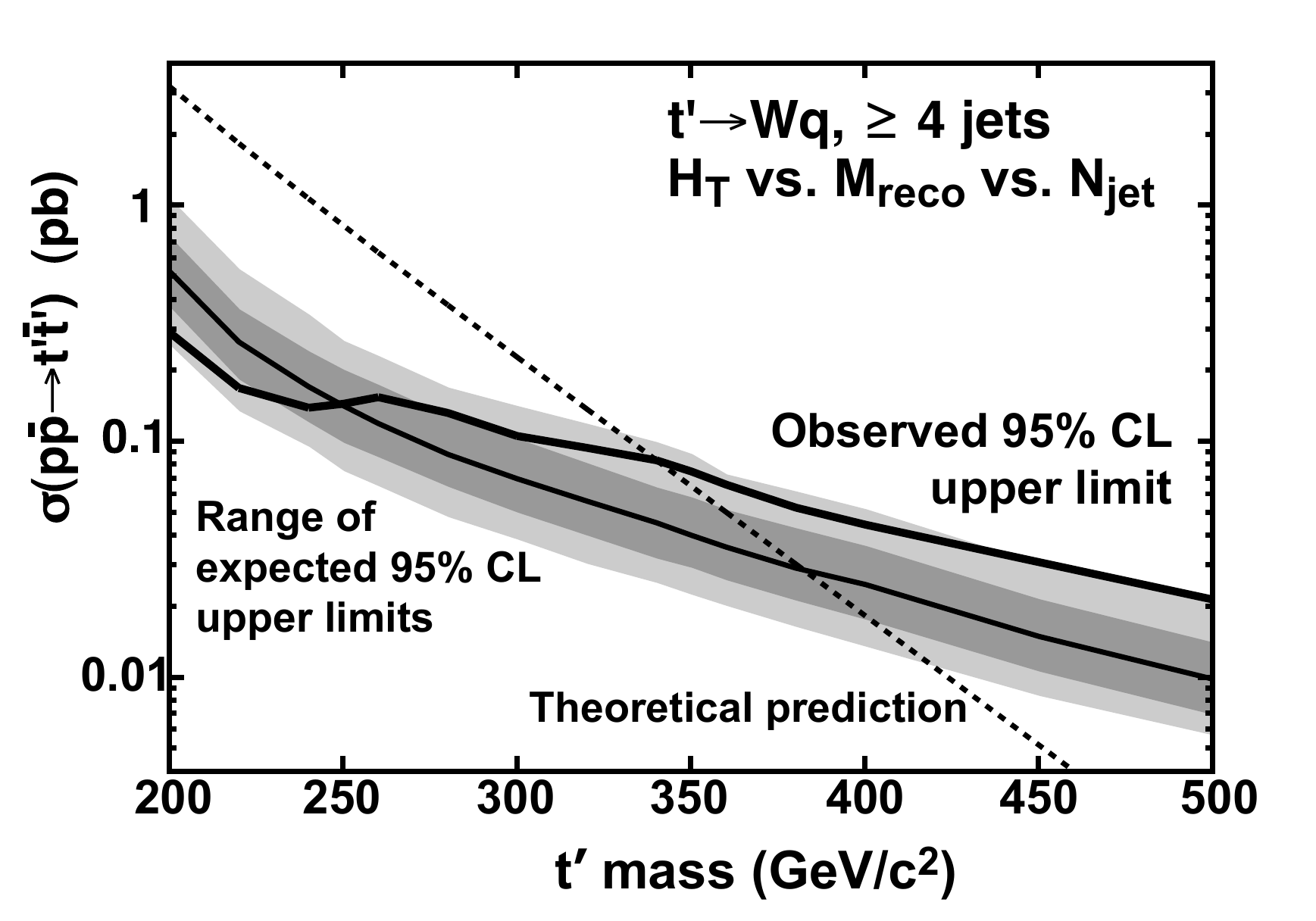}
          \caption{Observed and expected 95\% CL upper limits as a function of the mass of the $\tprime$ quark, for a $\tprime$ decaying to $Wb$ (upper) and $Wq$ (lower) with 100\% branching ratio. The light and dark gray areas show the $\pm$1$\sigma$ and $\pm$2$\sigma$ areas around the expected limits. The dashed line is the theory expectation.}
  \label{Figure:resultwb}
  \end{center}
\end{figure}

We test the sensitivity of our method by drawing pseudoexperiments from standard model distributions i.e., assuming 
no $\tprime$ contribution. The expected 95\% CL upper limits on the $\tprimepair$ production rate as a function of $\tprime$ mass, for a $\tprime$ decaying to Wb and Wq (assuming in either case a 100\% branching ratio) are shown in Fig.~\ref{Figure:resultwb}.
~The dashed line is the theoretical prediction for a fourth generation $\tprime$ with SM couplings~\cite{MMangano}.

We perform the analysis fit on the data which shows no significant excess from $\tprimepair$ production. Results expressed as a 95\% CL upper limit on the cross section are shown in Fig.~\ref{Figure:resultwb}. The individual limits along with the expected ones from pseudo-experiments are listed in Table~\ref{limit_curvewb} and \ref{limit_curvewq}.

\begin{figure}[htbp]
  \begin{center}
                \includegraphics[width=7.5cm,clip=]{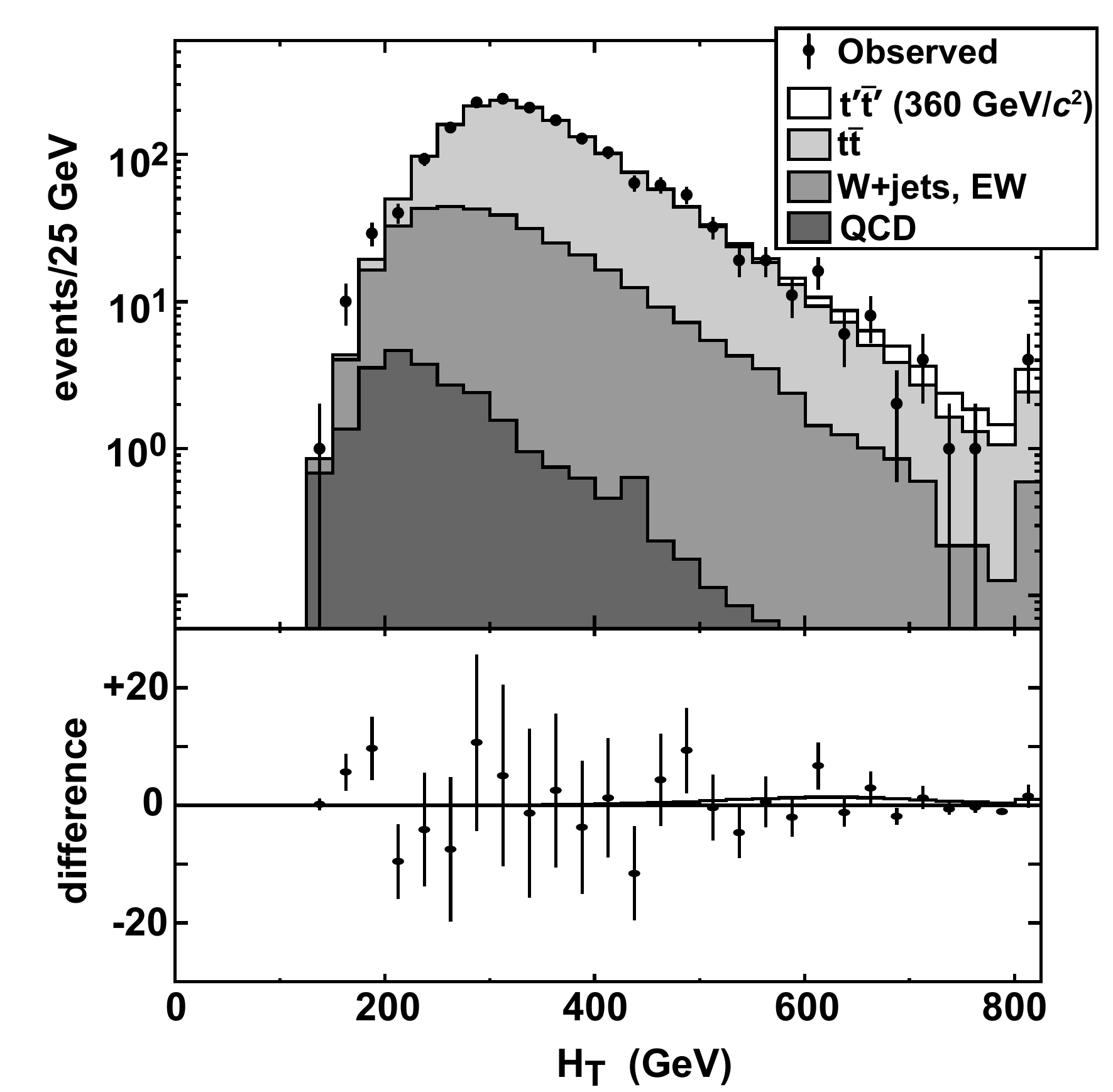} 
                     \includegraphics[width=7.5cm,clip=]{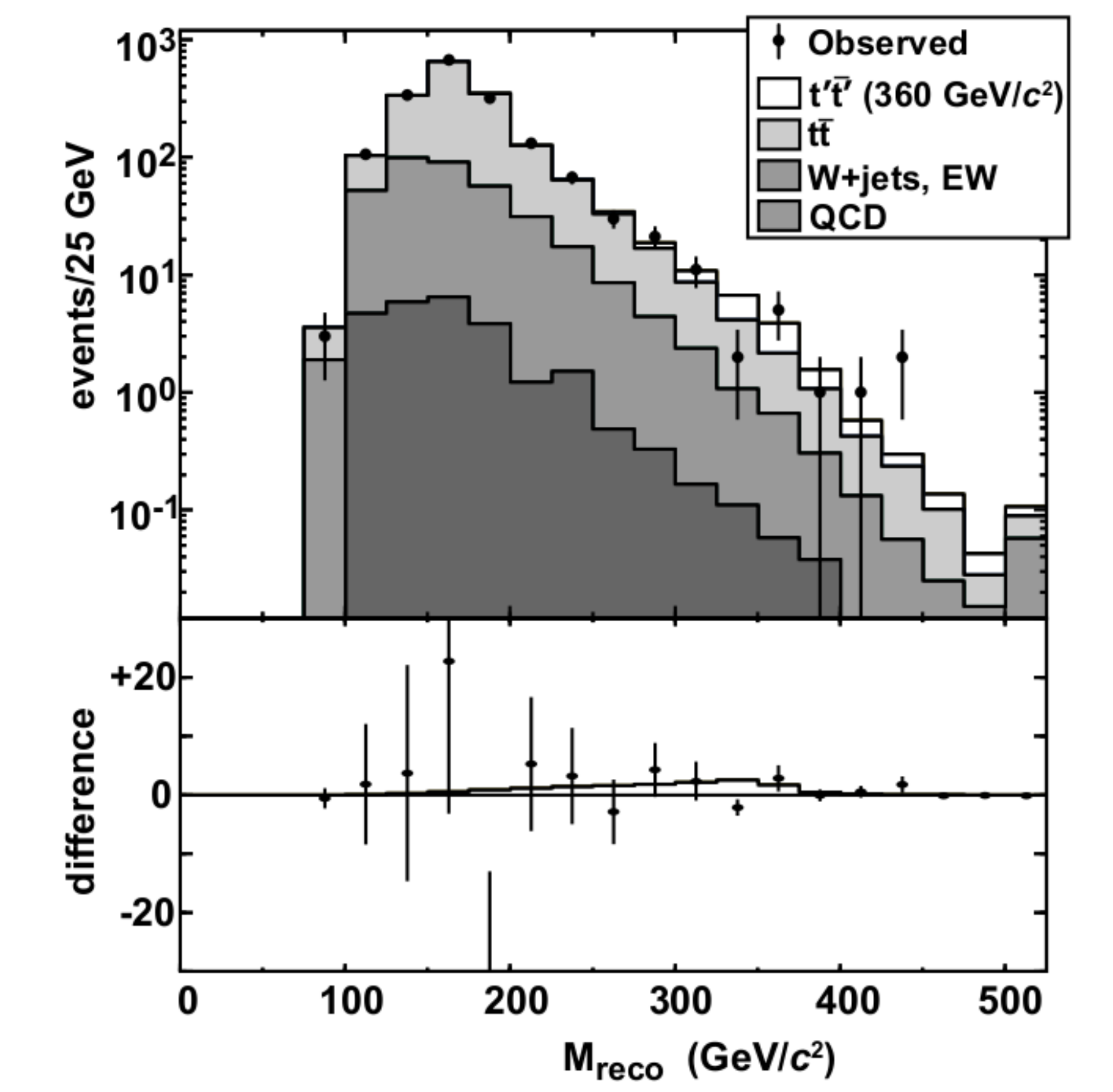} 
          \caption{Log scale distrubtions of $\sht$ and $\mrec$ comparing data (dots) with backgrounds (filled histograms) and signal (empty histogram). The $\tprimepair$ signal is for a $\tprime$ mass 360 GeV/$c^2$ and a $\tprimepair$ cross section corresponding to the 95\% CL upper limit. The amounts of all backgrounds are set to their fitted results from the fit assuming $\tprime$ decays to $Wb$. In the lower plot the points are the difference between the data and the sum of all the backgrounds, the histograms are the signal contribution. }
  \label{fig:all_sigwb}
  \end{center}
\end{figure}
\begin{figure}[htbp]
  \begin{center}
                \includegraphics[width=7.5cm,clip=]{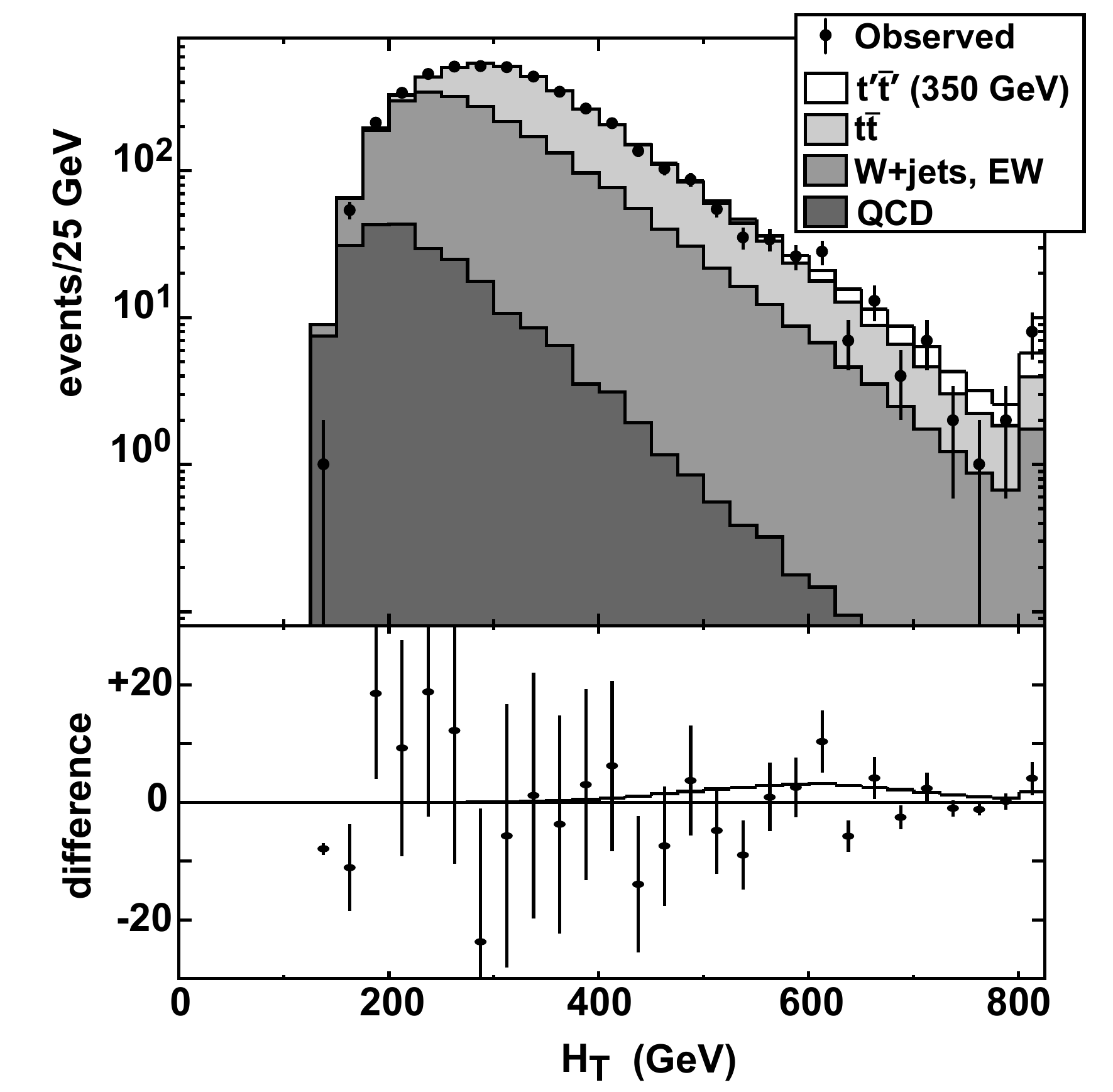} 
                     \includegraphics[width=7.5cm,clip=]{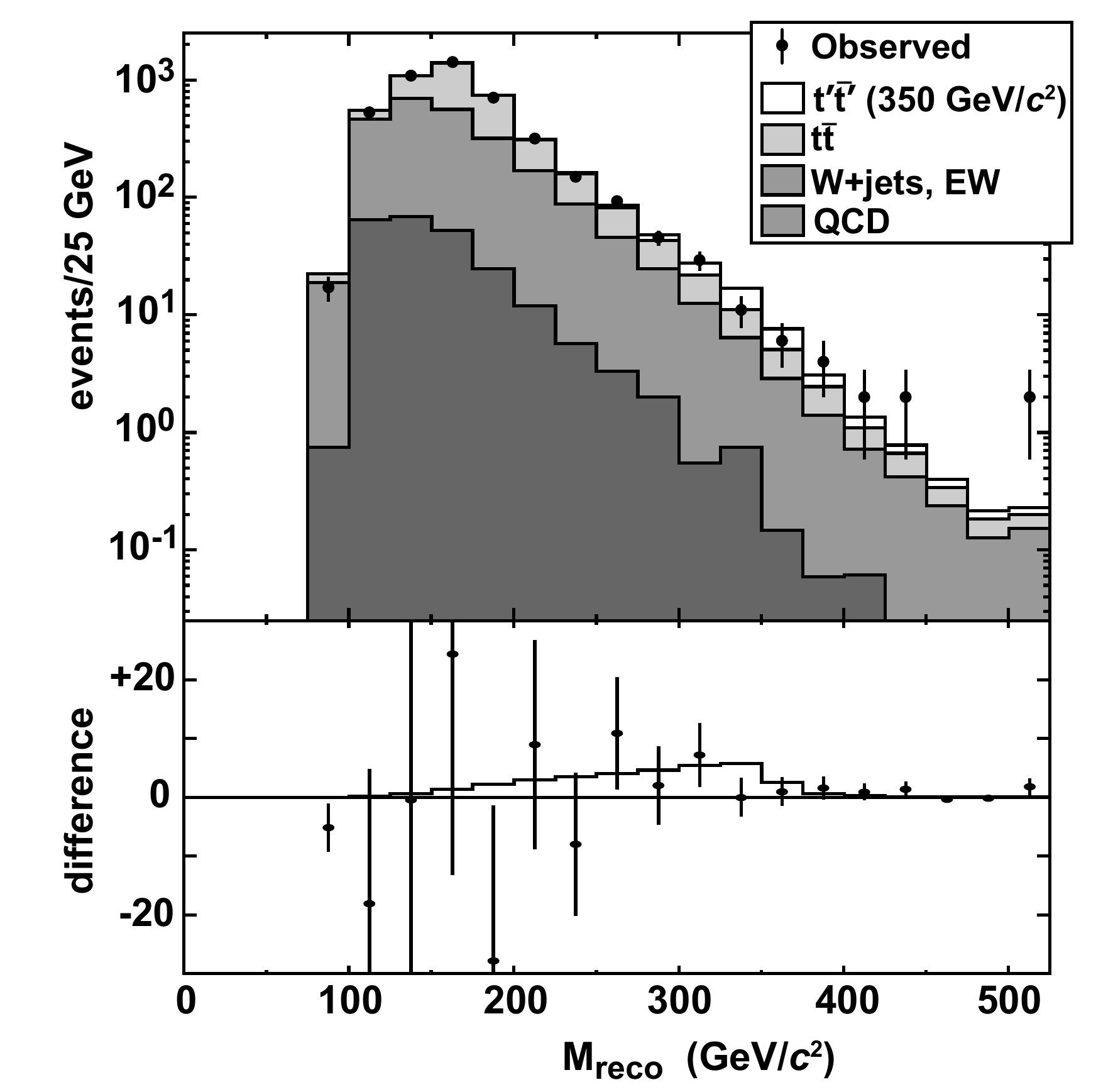} 
          \caption{Log scale distributions of $\sht$ and $\mrec$ comparing data (dots) with backgrounds (filled histograms) and signal (empty histogram). The $\tprimepair$ signal is for a $\tprime$ mass 350 GeV/$c^2$ and a $\tprimepair$ cross section corresponding to the 95\% CL upper limit. The amounts of all backgrounds are set to their fitted results from the fit assuming $\tprime$ decays to $Wq$. In the lower plot the points are the difference between the data and the sum of all the backgrounds, the histograms are the signal contribution. }
  \label{fig:all_sigwq}
  \end{center}
\end{figure}

Distributions of $\sht$ and $\mrec$ comparing the data with the fit to the backgrounds plus a signal contribution
are shown in Figs.~\ref{fig:all_sigwb} and \ref{fig:all_sigwq}. The backgrounds are normalized to their fitted results and the $\tprime$ signal with mass of 360 (350 for $\tprime \to Wq$) GeV/$c^2$ 
is normalized to its 95\% CL upper limit value.

In conclusion, we present a search for pair production of a $\tprime$ quark decaying to $Wq$, where q can be a generic down-type quark or specifically a $b$ quark. Having observed no excess attributable to $\tprimepair$ production, we exclude at 95\% CL a $\tprime$ quark with mass below 
358 (340) GeV/$c^2$ for $\tprime \to Wb (Wq)$. Examining the results separately for the cases where the $W$ decays to $e$ or $\mu$, we see no significant difference between them, obtaining separate limits of 292 (307 expected) GeV/$c^2$ for $\tprime \to Wb$ in the $\mu$ case and 306 (336 expected) GeV/$c^2$ for $\tprime \to Wb$ in the $e$ case. These are the most stringent limits set on such a quark at this time. While these direct limits are set on a fourth generation massive up-like 
quark $\tprime$, this analysis is sensitive to models of other massive quarks with similar signatures.

\linespread{1.4}{
\large
\begin{table}
  \begin{center}
\begin{tabular}{ccccc} \hline \hline

$m(\tprime)$ (GeV/$c^2$) & expected limit (pb) & observed limit (pb) \\ \hline
  180 &   1.757 $^{+0.729}_{-0.519}$  &   1.814 \\
 200  &   0.563 $^{+0.198}_{-0.178}$  &   0.581 \\
 220  &   0.209 $^{+0.099}_{-0.058}$  &   0.242 \\
 240  &  0.142 $^{+0.059}_{-0.041}$  &   0.139 \\
 250  &  0.121 $^{+0.047}_{-0.036}$  &   0.113 \\
 260  &  0.104 $^{+0.043}_{-0.029}$  &   0.106 \\
 280  &  0.082 $^{+0.034}_{-0.025}$  &   0.088 \\
 300  &   0.065 $^{+0.029}_{-0.018}$  &   0.076 \\ 
 320  &   0.052 $^{+0.023}_{-0.013}$  &   0.062 \\ 
 340  &  0.044 $^{+0.019}_{-0.011}$  &   0.057 \\ 
 350  &  0.040 $^{+0.019}_{-0.010}$   &   0.053 \\
 360  &  0.037 $^{+0.017}_{-0.010}$  &   0.054 \\ 
 380  &  0.032 $^{+0.013}_{-0.009}$  &   0.052 \\  
 400  &  0.028 $^{+0.011}_{-0.008}$   &   0.049 \\ 
 450  &   0.019 $^{+0.007}_{-0.006}$  &   0.031 \\ 
 500  &   0.013 $^{+0.006}_{-0.003}$  &   0.020 \\ \hline \hline
\end{tabular}
  \end{center}
  \caption{Expected, with $\pm1\sigma$ uncertainties, and observed limits on $\tprimepair$ production cross section for a given mass assuming the $\tprime$ quark decays to $Wb$.}
  \label{limit_curvewb}
\end{table}
}

\linespread{1.4}{
\large
\begin{table}
  \begin{center}
\begin{tabular}{ccc} \hline \hline

  $m(\tprime)$  (GeV/$c^2$)       & expected  limit (pb)    &       observed  limit (pb)   \\\hline
  180 &   1.116 $^{+0.506}_{-0.332}$  &   0.369 \\
 200  &   0.524 $^{+0.213}_{-0.153}$  &   0.290 \\
 220  &   0.263 $^{+0.100}_{-0.081}$  &   0.167 \\
 240  &  0.170 $^{+0.071}_{-0.050}$  &   0.138 \\
 250  &  0.141 $^{+0.060}_{-0.042}$  &   0.144 \\
 260  &  0.118 $^{+0.055}_{-0.032}$  &   0.153 \\
 280  &  0.088 $^{+0.039}_{-0.024}$  &   0.131 \\
 300  &   0.069 $^{+0.033}_{-0.019}$  &   0.105 \\ 
 320  &   0.056 $^{+0.025}_{-0.016}$  &   0.094 \\ 
 340  &  0.045 $^{+0.019}_{-0.013}$  &   0.083 \\ 
 350  &  0.040 $^{+0.019}_{-0.011}$   &   0.074 \\
 360  &  0.035 $^{+0.016}_{-0.009}$  &   0.065 \\ 
 380  &  0.029 $^{+0.014}_{-0.008}$  &   0.052 \\  
 400  &  0.025 $^{+0.011}_{-0.008}$   &   0.044 \\ 
 450  &   0.015 $^{+0.006}_{-0.004}$  &   0.031 \\ 
 500  &   0.010 $^{+0.004}_{-0.003}$  &   0.021 \\ \hline \hline
\end{tabular}
  \end{center}
  \caption{Expected, with $\pm1\sigma$ uncertainties, and observed limits on $\tprimepair$ production cross section for a given mass assuming the $\tprime$ quark decays to $Wq$.}
  \label{limit_curvewq}
\end{table}
}


We thank the Fermilab staff and the technical staffs of the participating institutions for their vital contributions. This work was supported by the U.S. Department of Energy and National Science Foundation; the Italian Istituto Nazionale di Fisica Nucleare; the Ministry of Education, Culture, Sports, Science and Technology of Japan; the Natural Sciences and Engineering Research Council of Canada; the National Science Council of the Republic of China; the Swiss National Science Foundation; the A.P. Sloan Foundation; the Bundesministerium f\"ur Bildung und Forschung, Germany; the World Class University Program, the National Research Foundation of Korea; the Science and Technology Facilities Council and the Royal Society, UK; the Institut National de Physique Nucleaire et Physique des Particules/CNRS; the Russian Foundation for Basic Research; the Ministerio de Ciencia e Innovaci\'{o}n, and Programa Consolider-Ingenio 2010, Spain; the Slovak R\&D Agency; and the Academy of Finland.




\begin{thebibliography}{99}
\bibitem{pseudo} CDF uses a cylindrical coordinate system with the $z$ axis along the proton beam axis. 
 				$\theta$ is the polar angle relative to the proton beam direction,
				and 	$\phi$ is the azimuthal angle.
				Missing transverse energy, $\met$, is defined as the magnitude of the vector $-\sum_{i}E^{i}_{T}\vec{n_i}$ 
				where $E^{i}_{T}$ are the magnitudes of transverse energy contained in each calorimeter tower $i$ and $\vec{n_i}$ 
				is the unit vector from the interaction vertex to the tower in the transverse $(x,y)$ plane.
				Pseudorapidity is  defined as $\eta \equiv -\ln(\tan\frac{\theta}{2})$, 
				 while transverse momenta and energies of particles are defined as 
                                      $p_T = |p|\sin\theta$ and $E_T = E\sin\theta$, respectively.
\bibitem{metsig} The significance of the missing transverse energy is defined as $\metsig = \frac{\met}{\sqrt{\sum_{jets}C^{2}_{JES}\cos^{2}
    (\Delta\phi_{\vec{\met},jet})+\cos^2(\Delta\phi_{\vec{\met}^{uncorr},\vec{\met}^{corr}})}}$ where $C_{JES}$ is a jet energy correction factor 
    and  $\Delta\phi_{\vec{\met}^{uncorr}, \vec{\met}^{corr}}$ is between the uncorrected and corrected $\vec{\met}$. 
    The $M_{T,W}$ for an event is defined as $M_{T,W} = \sqrt{2|p^{l}_{T}||p^{\nu}_{T}|(1-\cos(\Delta\phi(p^{l}_T,p^{\nu}_{T})))}$.
\end{thebibliography}
\end{document}